\documentclass{iopart}
\usepackage{graphicx,epsfig}
\usepackage{bm}
\usepackage{iopams}

\newcommand{\be}{\begin{equation}}
\newcommand{\ee}{\end{equation}}
\newcommand{\ba}{\begin{eqnarray}}
\newcommand{\ea}{\end{eqnarray}}
\newcommand{\bse}{\numparts}
\newcommand{\ese}{\endnumparts}

\newcommand{\bbq}{\begin{quote}}
\newcommand{\eeq}{\end{quote}}
\newcommand{\RR}{{}^3{\cal{R}}}

\newcommand{\HH}{{\cal{H}}}
\newcommand{\KK}{{\cal{K}}}
\newcommand{\PP}{{\cal{P}}}

\newcommand{\Omb}{\Omega_{b}}
\newcommand{\Ome}{\Omega_{e}}
\newcommand{\Omm}{\Omega_{m}}

\newcommand{\Ombi}{\Omega_{b0}}
\newcommand{\Omei}{\Omega_{e0}}
\newcommand{\Ommi}{\Omega_{m0}}

\newcommand{\Db}{\delta^{(b)}}
\newcommand{\Dj}{\delta^{(J)}}

\newcommand{\Dh}{\delta^{(\HH)}}
\newcommand{\dDh}{\dot\delta^{(\HH)}}

\newcommand{\Dm}{\delta^{(m)}}
\newcommand{\De}{\delta^{(e)}}

\newcommand{\Dk}{\delta^{(\kappa)}}

\newcommand{\dDrhob}{\dot\delta^{(b)}}
\newcommand{\dDrhom}{\dot\delta^{(m)}}
\newcommand{\dDrhoe}{\dot\delta^{(e)}}

\newcommand{\dd}{{\rm{d}}}

\begin{document}
\title{Evolution equations dynamical system of the Lemaître--Tolman--Bondi metric containing coupled dark energy.}
\author{Roberto C. Blanquet-Jaramillo${}^\dagger$, Roberto A. Sussman${}^\ddagger$, M\'aximo A. Ag\"{u}ero Granados${}^\dagger$, German Izquierdo${}^\dagger$}
\address{${}^\dagger$ Facultad de Ciencias, Universidad Aut\'onoma del Estado de M\'exico, Toluca 5000, Instituto literario 100, Edo. Mex.,M\'exico.\\
${}^\ddagger$ Instituto de Ciencias Nucleares, Universidad Nacional Aut\'onoma de M\'exico (ICN-UNAM),A. P. 70--543, 04510 M\'exico D. F., M\'exico.}
\ead{gizquierdos@uaemex.mx}
\date{\today}
\begin{abstract}We consider inhomogeneous spherically symmetric models based on the Lema\^{i}tre-Tolman-Bondi (LTB) metric, assuming as its source an interactive  mixture of ordinary baryonic matter, cold dark matter and dark energy with a coupling term proportional to the addition of energy densities of both dark fluids. We reduce Einstein's field equations to a first order 7-dimensional autonomous dynamical system of evolution equations and algebraic constraints. We study in detail the evolution of the energy density and spatial curvature profiles along the phase space by means of two subspace projections: a three-dimensional projection associated with the solutions of the Friedman-Lema\^\i tre-Robertson-Walker metric (invariant subspace) and a four-dimensional projection describing the evolution of the inhomogeneous fluctuations. We also classify and study the critical points of the system in comparison with previous work on similar sources, as well as solving numerically the equations for initial energy density and curvature profiles that lead to a spherical bounce whose collapsing time we estimate appropriately.\end{abstract}
\maketitle
\section{Introduction.}

Cosmological observations indicate that the Universe contains three primary matter-energy sources:  baryonic matter, cold dark matter (CDM) and dark energy (DE), respectively making 5 \%,\,27 \% and 68 \% of the total content \cite{planck}. Observational data strongly support the $\Lambda$CDM model in which dark energy is empirically described in our cosmic time by a cosmological constant equivalent to a source with pressure $p= -\rho$. However, this empiric model and other non-interactive dynamic dark energy models present a problematic ``coincidence problem'' that can be alleviated once we assume a non-trivial interaction between the dark sources (since it is safe to assume that non-gravitational interaction between the latter and visible matter must be very weak) \cite{caldera2009}. Several coupled models have been suggested and studied in detail (see examples in \cite{coupled-his}).

Observational data also supports a Friedman--Lema\^\i tre--Robertson--Walker (FLRW) background metric with energy density linear perturbations \cite{planck,copeland, wmap}, and observed local structure described by non--linear dynamics (whether Newtonian or relativistic) \cite{linperts}. Given the large amount of available DE and CDM models, it is necessary to contrast their predictions against observational data \cite{ol, ol2, Maar, Gavela}.

In order to study the structure formation, non--linear Newtonian dynamics is generally used (see review \cite{newt4}), as at larger scales (subhorizon) CDM can be well approximated by a pressure--less dust fluid, while a cosmological constant can play the role of the DE source. On the other hand, General Relativity is necessary to describe a more general DE source with a different pressure (whether interacting of not with the CDM, \cite{nonewt}). While numerical simulation involving continuous modeling or N--body solutions  can address the problem, inhomogeneous metrics that are exact solutions of Einstein's equations offer a idealized but interesting approach to the problem as they provide some analytical/physical results that complement the numerical work.

An example of the latter is the spherically symmetric Lema\^\i tre--Tolman--Bondi (LTB) metric. LTB metrics are typically associated in the astrophysical and cosmological literature with a pure dust source \cite{LTB}, with/without cosmological constant \cite{ltbrev1,ltbrev2,ltbrev3,ltbrev4}. However, the exact solutions provided by it are also compatible with nonzero pressures (something that it is often not known). In particular, it is possible to define a class of  ``quasi--local scalars'' (QL scalars \cite{RadAs,RadProfs,suss13a,suss13b}) that permit a clear, yet complete, description of the theoretical properties and evolution of sources with zero and nonzero pressure in terms of averages of standard covariant scalars satisfying FLRW dynamics, and the deviation from a FLRW background described by fluctuations with respect to the QL scalars \cite{suss13a}, all this with both the QL scalars and fluctuations being coordinate independent covariant quantities \cite{suss13b}. The QL scalars and their fluctuations transform Einstein's field equations for LTB models into evolution equations that can be set up as a self-consistent dynamical system (e.g., as has been done for dust models without and with a cosmological constant term, in \cite{suss08,sussmodes} and \cite{izsuss10} respectively). Connection of the LTB inhomogeneous metric with the FLRW linear perturbation theory is straightforward as a set of delta functions is defined (as the local scalar function divided by the corresponding QL quantity minus 1) that  can be put in correspondence with cosmological perturbations in the isochronous gauge \cite{sussmodes,suss15}.

LTB metrics are compatible with mixtures of an homogeneous DE fluid and an inhomogeneous CDM dust \cite{suss05}, but also with mixtures of dark fluids with anisotropic pressures \cite{sussQL,suss09} by means of the QL scalars associated with the isotropic pressure. We have studied previously LTB metric models by means of the QL formalism \cite{sussQL,suss08,izsuss10,suss15,izsuss17a,izsuss17b,sussPRD}, more recently considering as sources mixtures of non-relativistic CDM, described as dust, coupled to DE described as a dark fluid with constant equation of state $w<-1/3$. In \cite{izsuss17a} we assumed the coupling term to be proportional to CDM energy density, while in  \cite{izsuss17b} it was proportional to the DE density. In the present article, we generalize the previous results by considering a coupling term proportional to the addition  of both dark sources energy densities and considering an additional pressure--less uncoupled baryonic matter source. We study the 7--dimension dynamical system of the evolution equations and we classify the corresponding critical points. This analytical study give us an invaluable analytical information on the evolution of both QL scalars and perturbations that can help to understand and improve the numerical solutions. We also compute the evolution of a given set of initial conditions in order to illustrate the analytical findings. The initial profile chosen shows an scenario describing the outset of spherical collapse that could be interpreted as an idealized spherically symmetric structure formation example.

The section by section  disposition of the present article come next. In section \ref{genLTB} we describe the QL formalism and the corresponding differential equations for the LTB metric considered. In section \ref{dynsys}, we find the critical points in terms of the free parameters (FPs). In section \ref{numerical}, we study the necessary conditions to avoid singularities and set the initial profiles to illustrate an structure formation scenario. In section \ref{Interaction}, we compare the results found in this work with other coupling terms in the literature. Finally, in section \ref{conclusions}, we outline our findings. In this manuscript, we make use of natural units, $c=1$.

\section{LTB spacetimes, Q--scalar variables and coupled dark energy model}\label{genLTB}

The LTB metrics describe inhomogeneous spherically symmetric solutions that represent  exact local density perturbations tending asymptotically to an homogeneous FLRW metric. The models generalize the Newtonian spherically-symetric collapse in order to describe the evolution of non--relativistic spherical dust perturbations that start from a linear regime in the early Universe towards a fully non-linear regime just before virialization. This description allow us to consider a CDM and DE sources that provides a plain but useful generalization of the $\Lambda$-CDM model \cite{izsuss10, izsuss17a, izsuss17b}. The LTB metric can be written as %
\begin{equation}
ds^2 = -dt^2+\frac{R'^2\,dr^2}{1-K}+R^2[d\theta^2+\sin^2\theta\,d\phi^2],\label{LTB}
\end{equation}%
where $R=R(t,r)$ is a general function of the time $t$ and the radius coordinate $r$, $'=\partial /\partial r$, and $K=K(r)$ is a function related to the spatial curvature of the metric.

In order to model sources with non-trivial pressure we consider the most general energy--momentum tensor of the fluid compatible with the metric (\ref{LTB}) in a comoving frame with $u^a=\delta^a_t$
\begin{equation} T^{ab}=\rho\,u^a u^b+p\,h^{ab}+\Pi^{ab},\label{Tab} \end{equation}
where $\rho=\rho(t,r)$ y $p=p(t,r)$ are respectively the energy density and isotropic pressure and $\Pi^a_b=\PP(t,r)\times \hbox{diag}[0,-2,1,1]$ is the anisotropic pressure tensor of the fluid (a spacelike symmetric traceless tensor), while $h^{ab}=g^{ab}+u^au^b$ is the metric induced on the hypersurface at a constant time $t$.  Considering the fluid as a mixture of non-relativistic baryonic matter, together with non-relativistic CDM coupled to DE, the total energy density and isotropic pressure are
\ba
\rho=\rho_{b}+\rho_{m}+\rho_{e},\nonumber\\
p=p_{b}+p_{m}+p_{e},
\label{mixture}
\ea
where $\rho_{b}, p_b,\,\rho_{m}, p_m$ and $\rho_{e}, p_e$ are respectively the energy density and pressure of baryonic matter, CDM and DE mixture components. The conserved total energy--momentum tensor ($\nabla_b T^{ab}=0$) can be decomposed as
\begin{equation}
T^{ab}=T_{_{\rm{b}}}^{ab}+T_{_{\rm{m}}}^{ab}+T_{_{\rm{e}}}^{ab}.
\end{equation}
with an interaction between CDM and DE described by  energy-momentum flux (coupling current) between them as
\begin{equation} j^a=\nabla_b T_{(m)}^{ab}=-\nabla_b T_{(e)}^{ab}. \end{equation}
Given the symmetry of the metric (\ref{LTB}), $h_{ca}j^a=0$ holds as the current must be parallel to the 4--velocity. Then, $j_a=J u_a$ with  %
\begin{equation} J =u_a \nabla_b T_{m}^{ab}=- u_a \nabla_b T_{e}^{ab}, \label{uTab_cons_J} \end{equation}

Our LTB model, then, has seven local scalar fields dependent of the fluid surces whose evolution has to be solved: $A(t,r)=\rho_{b},\,\rho_{m},\,\rho_{e},\,p_{b},\,p_{m},\,p_{e}$, and $J$. Following the quasi--local scalars (QL scalars) description of the LTB metrics \cite{izsuss17a,izsuss17b,sussQL,suss09}, QL scalar $A_q$ and fluctuation $\delta^A$ are defined for every $A$ as%
\begin{equation} A_q = \frac{\int_{x=0}^{x=r}{A\,R^2 R' dx}}{\int_{x=0}^{x=r}{R^2 R' dx}} \label{QLfunc},\qquad\delta^A=\frac{A-A_q}{A_q}=\frac{A'_q/A_q}{3R'/R}, \label{qmaps} \end{equation} %
where $x=0$ is a symmetry centre of the metric and $R(t,0)=\dot{R}(t,0)=0$, with  $\dot{}=\partial /\partial t$, and $\dot{R}=u^a\nabla_a R$. The QL pressures of the sources are related to the anisotropic pressures as follows %
\ba
p_{bq}&=&p_{b}-2\PP_{b},\quad \delta_{(b)}^p=2\PP^{(b)},\nonumber\\
p_{mq}&=&p_{m}-2\PP^{(m)} \quad \delta_{(m)}^p=2\PP^{(m)},\\
p_{eq}&=&p_{e}-2\PP^{(e)},\quad \delta_{(e)}^p=2\PP^{(e)}.\nonumber \label{pps}
\ea%

Additionally, there are two covariant scalars associated to the metric (\ref{LTB}): the Hubble expansion scalar $\HH=(1/3)\nabla_a u^a=(R^2R')\,\dot{}/(R^2R')$, and the spatial curvature $\KK=(1/6)\RR=2(KR)'/(R^2R')$, where the later is related to the Ricci scalar $\RR$ of constant $t$ hypersurfaces with induced metric $h_{ab}$. The corresponding QL scalars read
\begin{equation} \HH_q=\frac{\dot R}{R},\qquad \KK_q=\frac{K}{R^2}. \label{HKq} \end{equation}
The local interaction term $J$ is also scalar and defines a QL interaction $J_q$ and a corresponding delta ($J=J_q(1+\Dj)$). In this work, we consider that $J_q$ depends on the rest of QL scalars and will be defined later. In order to obtain the evolution equations of the model, we need to consider the following equations of state (EOS) for the different sources
\ba
 \hbox{baryonic matter (dust):}\qquad\qquad\quad p_{b}=0\quad\Rightarrow\quad \delta_{(b)}^p=0 \label{eosbaryonic} \\
 \hbox{CDM (dust):}\qquad\qquad\quad p_{m}=0 \quad\Rightarrow\quad \delta_{(m)}^p=0, \label{eoscdm} \\
 \hbox{DE (barotropic fluid):}\qquad p_{e}=w\rho_{e} \quad\Rightarrow\quad \delta_{(e)}^p=\delta_{(e)}^\rho,\label{eosde}
\ea
where we have assumed that $w$ is a constant. According to equations (\ref{eosbaryonic}-\ref{eosde}), the DE source is the only one that contributes to the anisotropic pressure: $\PP=\PP^{(e)}=\delta_{(e)}^p/3$. As in \cite{izsuss17a,izsuss17b,sussQL,suss09}, the evolution equations are %

\bse\label{eveqs_q2}\ba \dot{\HH}_q&=&-\HH_q^2-\frac{\kappa}{6}\,\left[\rho_{mq}+(1+3\,w\,)\rho_{eq}+\rho_{bq} \right]\label{evHH_q2}\\
\dot{\rho}_{bq} &=& -3\HH_{q}\,\rho_{bq},\label{evb_q2}\\
\dot{\rho}_{mq} &=& -3\HH_q\,\rho_{mq}+J_{q},\label{evm_q2}\\
\dot{\rho}_{eq} &=& -3\HH_{q}\left( 1+w \right)\,\rho_{eq}-J_{q},\label{eve_q2}\\
\dDh &=& -\HH_q\Dh\left(1+3\Dh\right)+\frac{\kappa}{6\HH_q}\left[\rho_{mq}\,\left(\Dh-\Dm\right)\right.\nonumber\\
        &&\left.+(1+3w)\rho_{eq}\,\left(\Dh-\De\right)+\rho_{bq}\,\left(\Dh-\Db\right)\right],\label{evDh_q2}\\
\dDrhob&=& -3\HH_q\,\Dh\left[1+\Db\right],\label{evDb_q2}\\
\dDrhom&=&-3\HH_q\,\left(1+\Dm\right)\Dh+\frac{J_q}{\rho_{mq}}\left(\Dj-\Dm\right),\label{evDm_q2}\\
\dDrhoe&=&-3\HH_q\,\left(1+w+\De\right)\,\Dh-\frac{J_q}{\rho_{eq}}\left(\Dj-\De\right)\label{evDe_q2}. \ea\ese%
with the constraints %
\ba \HH_q^2&=&\frac{\kappa}{3}\,\left[\rho_{bq}+\rho_{mq}+\rho_{eq}\right]-\KK_q, \label{cHam2}\\
2\HH_q^2\Dh&=&\frac{\kappa}{3}\left(\rho_{bq}\Db+\rho_{mq}\Dm+\rho_{eq}\De\right)-\KK_q\Dk. \label{cHam2b} \ea
where $\kappa=8\pi G$, (\ref{cHam2b}) follows from (\ref{cHam2}) and (\ref{qmaps}), $\Dk=(\KK-\KK_q)/\KK_q$. The evolution equations (\ref{evHH_q2}--\ref{eve_q2})and the constraint(\ref{cHam2}), at every comoving shell $r=r_i$, are similar to the corresponding FLRW evolution equations.

Also, in the limit $r\to\infty$, the LTB metric can be matched to a FLRW background as $\Db,\,\Dm,\,\De,\,\Dh,\,\Dj$ vanish, \cite{RadAs}. The differential equations system (\ref{evHH_q2}--\ref{evDe_q2}) depends on the FP $w$ and the scalar $J_q$. In this work, the $J_q$ considered is
\ba
J_{q}&=&3\HH_{q}\,\alpha\,\left( \rho_{mq}+\rho_{eq} \right) \label{intj_3me}\\
\Dj&=&\Dh+\frac{\rho_{mq}\,\Dm}{\left( \rho_{mq}+\rho_{eq} \right)}+\frac{\rho_{eq}\,\De}{\left( \rho_{mq}+\rho_{eq} \right)},\label{djint3}
\ea
where $\alpha$ is a dimensionless coupling constant. This coupling term is considered in the literature (see \cite{ol,ol2, ol3, wang16}) in the context of FLRW cosmology. It represents a generalization of the coupling terms used in \cite{izsuss17a, izsuss17b}, $J_{q}=3\HH_{q}\alpha \rho_{mq}$ and $J_{q}=3\HH_{q}\alpha \rho_{eq}$, respectively. Given the evolution of the QL energy densities of CDM and DE (CDM dominating the early universe expansion, while DE dominates the late expansion), it is expected that the coupling (\ref{intj_3me}) behaves as the coupling in \cite{izsuss17a} in the early universe and behaves as the coupling in \cite{izsuss17b} for the late expansion of the metric.

For $\alpha>0$ and with the definition in (\ref{uTab_cons_J}),  it is straightforward to conclude that energy density flows from the DE to the CDM. This coupled DE model have been studied in the frame of FLRW metrics in \cite{copeland,ol,Maar}, where the energy density flux term ($Q$) is an homogeneous scalar that represents phenomenologically the microscopical interaction between the DE scalar field and the CDM particles.

\subsection{QL scalars scaling laws}\label{scalinglaws}

Metric (\ref{LTB}) can be rewritten as
\ba \dd s^2 = -\dd t^2 +L^2\,\left[\frac{\Gamma^2\,R'_0{}^2 \dd r^2}{1-\KK_{q0} R_{0}^2 }+R_{0}^2\,(\dd\theta^2+\sin^2\theta\dd\phi^2)\right],\label{LTB2}\\
\Gamma =\frac{{R'/R}}{R'_0/R_0} =1+ \frac{L'/L}{R'_0/R_0},\ea
where $L=L(t,r)$ is a generalization of the FLRW scale factor. Given the LTB metric invariance under radial coordinate rescaling, it is possible to  specify the function $R_0(r)$ in order to define a physical radial coordinate $\mathfrak{R}=R_0(r)$, as $d\mathfrak{R}=R'_0 dr$. We can set the Big Bang singularity at the instant for which $L(t,r)=0$, while $\Gamma(t,r)=0$ would define a shell crossing singularity \cite{izsuss10}.

Solving evolution equations (\ref{HKq}), (\ref{evHH_q2}), (\ref{evb_q2}),(\ref{evm_q2}) and (\ref{eve_q2}) with respect to $L$, we obtain the scaling laws for the  QL scalars (equivalent to scaling laws of the analogous FLRW scalars),
\ba
\KK_q&=&\KK_{q0}L^{-2},\quad
\rho_{bq} =\rho_{bq0}L^{-3},\nonumber\\
\rho_{eq} &=&\rho_{eq0}\left(a L^{\gamma_1}+(1-a) L^{\gamma_2}\right)+\rho_{mq0}b(-L^{\gamma_1}+L^{\gamma_2}),\quad \nonumber \\
\rho_{mq} &=&\rho_{mq0}\left((1-a)L^{\gamma_1}+a L^{\gamma_2}\right)+\rho_{eq0}b(L^{\gamma_1}-L^{\gamma_2}),
\label{scalinglawrhomq}
\ea
where
\ba
a&=&\frac{1}{2}+\frac{2\alpha+w}{2w\Delta},\, b=-\frac{\alpha}{w\Delta},\nonumber\\
\gamma_1&=&-\frac{3}{2}(2+w(1+\Delta)),\, \gamma_2=-\frac{3}{2}(2+w(1-\Delta))\nonumber,\\
\Delta&=&\sqrt{1+4(\alpha/w)}.
\ea

\section{The non-dimensional evolution equations system}\label{dynsys}

In order to transform (\ref{evHH_q2}--\ref{evDe_q2}) into a proper autonomous dynamical system associated with cosmological variables, it is necessary to define the following dimensionless energy density $\Omega$ functions
\begin{equation} \Omb=\frac{\kappa}{3\HH_q^2}\rho_{bq},\qquad \Omm=\frac{\kappa}{3\HH_q^2}\rho_{mq},\qquad \Ome=\frac{\kappa}{3\HH_q^2}\rho_{eq}.\label{omegas} \end{equation}
whose evolution equations follow from (\ref{evHH_q2}), (\ref{evb_q2}), (\ref{evm_q2}), and (\ref{eve_q2})
as
\begin{equation} \frac{1}{\HH_q}\dot{\Omega}_A=\frac{\kappa}{3\HH_q^3}\dot{\rho}_{Aq}-{\Omega}_A\frac{\dot{\HH_q}}{\HH_q}, \end{equation}%
with $A=b,m,e$. The constraints (\ref{cHam2}) and (\ref{cHam2b}) in terms of the $\Omega$ functions are
\ba
\Omb+\Omm+\Ome+{\Omega}_{\KK}=1, \label{cHam3a} \\
2\Dh=\Omb\Db+\Omm\Dm+\Ome\De+{\Omega}_{\KK}\Dk. \label{cHam3b} \ea
where ${\Omega}_{\KK}=-\KK_q/\HH^2_q$. It is convenient to define for all comoving observers $r=r_i$ a dimensionless time coordinate $\xi(t,r)$ \cite{izsuss10}
\begin{equation} \frac{\partial}{\partial\xi}=\frac{1}{\HH_q}\frac{\partial}{\partial t}=\frac{3}{\Theta_q}\frac{\partial}{\partial t}. \label{xidef} \end{equation}
where we remark that surfaces of constant $\xi$ do not (in general) coincide with surfaces of constant $t$ (they can coincide only for a given initial fixed $t=t_i$ identified with an initial $\xi_i$).

In terms of $\xi$, and using (\ref{intj_3me}) and (\ref{djint3}), the system (\ref{evHH_q2}--\ref{evDe_q2}) becomes
\bse\label{eveqs_q3}\ba \frac{\partial{\Omb}}{\partial{\xi}} &=& \Omb\,\left[ -1+\Omm+(1+3\,w)\,\Ome+\Omb \right],\label{evsistemdinamic3a}\\
\frac{\partial{\Omm}}{\partial{\xi}} &=& \Omm\,\left[ -1+3\,\alpha+\Omm+(1+3\,w)\,\Ome+\Omb \right]+3\,\alpha\Ome,\label{evsistemdinamic3b}\\
\frac{\partial{\Ome}}{\partial{\xi}} &=& \Ome\,\left[ -1-3\,w-3\,\alpha+\Omm+(1+3\,w)\,\Ome+\Omb\right]-3\,\alpha\Omm,\label{evsistemdinamic3c}\\
\frac{\partial{\Dh}}{\partial{\xi}} &=& -\Dh\left(1+3\Dh\right)+\frac{\Omm\,\left(\Dh-\Dm\right)}{2}\nonumber\\
    &&+\frac{(1+3w)\Ome\,\left(\Dh-\De\right)}{2}+\frac{\Omb\,\left(\Dh-\Db\right)}{2},\label{evsistemdinamic3d}\\
\frac{\partial{\Db}}{\partial{\xi}} &=& -3\Dh\left[1+\Db\right],\label{evsistemdinamic3e}\\
\frac{\partial{\Dm}}{\partial{\xi}} &=& -3\,\left(1+\Dm\right)\,\Dh+3\,\alpha\,\Dm-\frac{3\,\alpha\,\Dm\,\left(\Omm+\Ome \right)}{\Omm}\nonumber\\
    &&+\frac{3\,\alpha\,\Dh\,\left(\Omm+\Ome \right)}{\Omm}+\frac{3\,\alpha\,\De\,\Ome}{\Omm},\label{evsistemdinamic3f}\\
\frac{\partial{\De}}{\partial{\xi}} &=& -3\,\left(1+w+\De\right)\,\Dh-3\,\alpha\,\De+\frac{3\,\alpha\,\De\,\left(\Omm+\Ome \right)}{\Ome}\nonumber\\
&&-\frac{3\,\alpha\,\Dh\,\left(\Omm+\Ome \right)}{\Ome}-\frac{3\,\alpha\,\Dm\,\Omm}{\Ome}.\label{evsistemdinamic3g} \ea\ese
The autonomous dynamical system (\ref{evsistemdinamic3a}--\ref{evsistemdinamic3g}) has seven-dimensions and can be numerically solved for initial conditions given at a fixed $\xi=\xi_i$ for each comoving shell $r=r_i$.

The density variables $\Omb$, $\Omm $ and $\Ome$ form a separate subsystem (as eqs. (\ref{evsistemdinamic3a}--\ref{evsistemdinamic3c}) do not depend on the $\delta$ functions). Hence, this subsystem is an invariant subspace of (\ref{evsistemdinamic3a}--\ref{evsistemdinamic3g}), formally identical to the dynamical system that would be obtained for $\Omega$ functions of the corresponding FLRW model with null $\delta$ functions. We will refer to the set (\ref{evsistemdinamic3a}--\ref{evsistemdinamic3c}) as the homogeneous projection (subsystem). On the other hand, the $\delta$ functions depend on both $\delta$ and $\Omega$ functions, and only form an independent subsystem when the $\Omega$ are constant during the evolution (i.e., when $\Omega$ functions do not evolve at the critical points of the homogeneous projection). We will refer to the set of $\delta$ functions evolution equations (\ref{evsistemdinamic3d}--\ref{evsistemdinamic3g}) as the inhomogeneous projection. It is possible to fully represent the solution for a set of initial conditions at a given shell $r=r_i$ by means of a trajectory evolution  3-dimensional plot in the homogeneous phase-space plus the evolution of the $\delta$ functions vs. $\xi$ (or vs. $t$).

The QL scalar $\HH_q(\xi,r_i)$ is related to the $\Omega$ functions as\footnote{In fact, for the numerical work it is convenient to add equation (\ref{HOm}) to the system (\ref{evsistemdinamic3d}--\ref{evsistemdinamic3g}), and solve the resulting 8--dimensions system for a given set of initial conditions related by the constraints.}
\be
\frac{\partial{\HH_q}}{\partial{\xi}}=\frac{\dot{\HH}_q}{\HH_q}=-\HH_q\left(1+\frac{1}{2}\Omb+\frac{1}{2}\Omm+\frac{1+3w}{2}\Ome\right).\label{HOm}
\ee
From the numerical solutions $\Omb,\,\Omm,\,\Ome,\,\Db,\,\Dm,\,\De,\,\Dh$ and $\HH_q(\xi,r_i)$, it is possible to compute the rest of the scalars that characterize the LTB metric: the QL baryonic, CDM and DE densities from (\ref{omegas}), the spatial curvature and its fluctuation $\Dk$ from the constraints (\ref{cHam3b}), and the corresponding local scalars $A=\HH,\,\KK,\,\rho_b,\,\,\rho_m,\rho_e,\,J$ from $A=A_q(1+\delta^A)$.

The cosmic physical time can be computed as well at a fixed $\xi(t ,r)$ and $r=r_i$ \cite{izsuss10}
\be
t(r_i)= \int_0^{\xi(t,r_i)} {\frac{d \xi'}{\HH_q(\xi',r_i)}}.\label{phystime}
\ee
And finally, the scalars appearing in LTB metric can be computed as $R=\exp\left(\int{\HH_q dt}\right)$ and $R'=R\exp\left(-\int{\HH_q \Dh dt}\right)$.

\subsection{Homogeneous subspace}

We obtain the critical points of (\ref{evsistemdinamic3a}--\ref{evsistemdinamic3c}) by solving the algebraic quadratic equations that follows by setting to zero their right--hand side. We also compute the eigen--value set related to each critical point by linearization of the homogeneous system  near the critical points, by means of the jacobian matrix of the dynamical system. The findings are summarized in table \ref{crit-points}.
\begin{table}[tbp]\label{table1}
\centering\caption{The critical points and their respective eigenvalues of the system (\ref{evsistemdinamic3a}--\ref{evsistemdinamic3c}).}\label{crit-points}
\begin{tabular}{|l|l|l|}\hline\begin{tabular}[c]{@{}l@{}}Critical points \end{tabular}
& $(\Omb,\Omm,\Ome)$ &
Eigenvalues ($\Delta=\sqrt{1+4(\alpha/w)}$) \\ \hline
P1 & $(0,\frac{\left(1+\Delta\right)}{2},\,\frac{\left(1-\Delta\right)}{2})$ & \begin{tabular}
[c]{@{}l@{}}
$\lambda_1=-3w\Delta,\,\lambda_2=\frac{3w}{2}(1-\Delta),\,\lambda_3=1+\frac{3w}{2}(1-\Delta)$\\ \end{tabular} \\ \hline

P2 &$(0,\,\frac{\left(1-\Delta\right)}{2},\,\frac{\left(1+\Delta\right)}{2})$ &\begin{tabular}[c]{@{}l@{}}
$\lambda_1=\frac{3w}{2}(1+\Delta),\, \lambda_2=1+\frac{3w}{2}(1+\Delta),\, \lambda_3=3\,w\Delta$\\ \end{tabular} \\ \hline

P3 & $(1,\,0,\,0)$ & \begin{tabular}[c]{@{}l@{}}
$\lambda_1=-\frac{3w}{2}(1-\Delta),\,\lambda_2=-\frac{3w}{2}(1+\Delta),\,\lambda_3=1$\\ \end{tabular} \\ \hline
P4 & $(0,\,0,\,0)$ &\begin{tabular}[c]{@{}l@{}}
$\lambda_1=-1-\frac{3w}{2}(1-\Delta),\,\lambda_2=-1-\frac{3w}{2}(1+\Delta),\,\lambda_3=-1$ \end{tabular} \\ \hline \end{tabular} \end{table}

Points $P1$, and $P2$ could be mathematically correct but non--physical (complex solutions or negative defined real numbers), for general values of $\alpha$ and $w$. As $w<-1/3$ for FLRW DE models (DE being the responsible of the late accelerated expansion stage of the universe), we see that the points are real for $\alpha \leqslant -w/4$. This result is discussed in \cite{ol3} in the context of FLRW cosmology. Considering the FPs for the interaction that fulfills the observational bounds obtained in \cite{ol,ol2,ol3,wang16}, we see that both points are always physical when $w$ is close to/lower than $-1$, and $\alpha<0.25$.

Also, it is possible to obtain non--negative $P1$ and $P2$ when $\alpha>0$. While \cite{izsuss17a} considers also the evolution of the metric (\ref{LTB}) with $\alpha<0$ (for a different coupling term), in the present article we will restrict ourselves to $\alpha>0$ in order to study the relevance of the critical points $P1$ and $P2$ on the dynamics of the metric (\ref{LTB}). The study of the phase space evolution will be undertaken by looking at the trajectories in terms of the corresponding homogeneous projection.

The critical points and their respective eigen--values are displayed in Table (\ref{crit-points}). Considering  $w\sim-1$ and $0<\alpha\leq -w/4$, it is possible to determine the behavior of the trajectories near the critical points. The critical point $P1$ is a saddle point as $\lambda_1>0$ while the rest of the eigenvalues are strictly negative. The critical point $P2$ is a future attractor (with negative eigenvalues). $P3$ is a past attractor with all eigenvalues strictly positive. Finally, $P4$ is a saddle point as $\lambda_1, \, >0$ while $\lambda_2,\lambda_3<0$.

The points of the homogeneous subspace that live on the plane $M\equiv\Omb+\Omm+\Ome=1$ (and from (\ref{cHam3a}),  $M\equiv{\Omega}_{\KK}=0$) form an invariant subspace. Defining a vectorial base of the homogeneous subsystem $\{\bm{u_1},\bm{u_2},\bm{u_3}\}$ as the orthonormal vectors in the direction of the $\Omb$, $\Omm$, and $\Ome$ axis, respectively, it is possible to define a new base $\{\bm{v_1},\bm{v_1},\bm{n}\}$, where $\bm{v_1}, \bm{v_1}$ are two linear independent vectors generating the invariant plane while $\bm{n}=(\bm{u_1}+\bm{u_2}+\bm{u_3})/\sqrt{3}$ is the normal vector to the plane. Any trajectory curve in the phase space is written as
\ba
\bm{\Omega}(\xi)&=&\Omb(\xi)\bm{u_1}+\Omm(\xi)\bm{u_2}+\Ome(\xi)\bm{u_3}\nonumber\\
&=&\Omega_1(\xi)\bm{v_1}+\Omega_2(\xi)\bm{v_1}+\Omega_n(\xi)\bm{n}
\ea
where $\Omega_i(\xi)=\bm{\Omega}\cdot\bm{v_i}$ with $i=1,2$ and $\Omega_n(\xi)=\bm{\Omega}\cdot\bm{n}=(\Omb+\Omm+\Ome)/\sqrt{3}$. We can solve the evolution of the trajectories in the direction of the second base from eqs. (\ref{evsistemdinamic3a}-\ref{evsistemdinamic3c}). In particular,
\ba
\frac{d}{d \xi}\Omega_n&=&\frac{1}{\sqrt{3}}\left(\frac{d\Omb}{d \xi}+\frac{d\Omm}{d \xi}+\frac{d\Ome}{d \xi}\right)\nonumber\\
&=&\frac{1}{\sqrt{3}}\left((\Omb+\Omm+\Ome-1)(\Omb+\Omm+\Ome+3w\Ome)\right).
\ea
From the relation of above it is clear that the trajectories with $\Omb+\Omm+\Ome-1=-{\Omega}_{\KK}=0$ at any point do not evolve in the direction of $\bm{n}$, and, consequently, live in the invariant plane entirely. , as for any point $P\in M$

\be
\left[\frac{d}{d \xi}\left(\Omb+\Omm+\Ome\right)\right]_P=0.
\ee
The trajectories in the homogeneous phase space cannot cross the invariant subspace $M$ \cite{izsuss17a,izsuss17b}, so they maintain the same ${\Omega}_{\KK}$ sign during their entire evolution. The homogeneous subspace is divided in three separate regions: trajectories for which ${\Omega}_{\KK}=0$, trajectories with ${\Omega}_{\KK}>0$, and trajectories with ${\Omega}_{\KK}<0$.

Given that the critical points P1, P2, and P3 are on the invariant plane (P2 and P3 being future and past attractors, respectively), it seems contradictory for the trajectories with non null curvature to evolve to/from them. At this point, it is important to stress that the trajectory evolution to the point P2, independently of the curvature, is asymptotical, i.e., in the limit $L\rightarrow \infty$ in the expanding LTB solution. From (\ref{scalinglawrhomq}), we know $\KK_q$, $\rho_{bq}$ evolve as $L^{-2}$ and $L^{-3}$, respectively, while both $\rho_{mq}$ and $\rho_{eq}$ evolve as an addition of a term with $L^{\gamma_1}$ plus a term with $L^{\gamma_2}$. When $L\rightarrow \infty$, $\KK_q$, $\rho_{bq}$ will become much smaller than $\rho_{mq}$ and $\rho_{eq}$ as $\gamma_1>-2$ for the FPs considered. The same discussion applies to the past attractor $P3$ as $L\rightarrow 0$, the baryonic matter $\rho_{bq}$ has asymptotical values greater than those of $\KK_q$, $\rho_{mq}$, and $\rho_{eq}$ in the past and is the dominant source ($-3<\gamma_2$). On the other hand, considering an additional uncoupled radiation source, as in \cite{izsuss17b}, we would find that the point P3 is no longer a past attractor but a saddle point and a new past attractor appears where only the radiation source is non null (with QL energy density scaling as $L^{-4}$). It is convenient to consider the radiation source in order to obtain the standard Cosmology radiation expansion stage.

\subsection{Complete dynamical system critical points}

Fixing the $\Omega$ functions to the values given by the homogeneous critical points, we now find the inhomogeneous part of them by solving the $\delta$ functions values that make the right hand side of eqs. (\ref{evsistemdinamic3d}-\ref{evsistemdinamic3g}) null. We also study the behaviour of the system in the vicinity of the points by finding the eigen-values of the jacobian matrix of the complete system at the points. Some mathematical solutions for the critical points include $\delta$ functions lower than $-1$ for some choices of parameters, we will consider that these solutions are not compatible with the spherical symmetry. As discussed in \cite{izsuss17a,izsuss17b}, a $\delta$ function evolving to values lower than $-1$ could be interpreted as a break in the spherical symmetry that should be addressed with a more general metric, such as the non--spherical Szekeres metric. For the critical points of the homogeneous subspace P3 and P4, $\Omm$ and $\Ome$ are null and the right hand side of eqs. (\ref{evsistemdinamic3f},\ref{evsistemdinamic3g}) cannot be evaluated as some terms depend on the ratio $\Omm/\Ome$, or its inverse. However, in the vicinity of the points P3 and P4, we will study the corresponding eigen-values of the jacobian matrix of the complete system in the limit
\be
lim_{(\Omm,\Ome)\rightarrow(0,0)}\frac{\Ome}{\Omm}=k.
\ee
The limit of above depends strongly on the curve $\Ome=\Ome(\Omm)$ considered in the plane $\Ome-\Omm$. If we assume a curve such that the limit is a non-null constant $k$, it is possible to find the eigen--values of the critical points as a function of the FPs and also the direction constant $k$. 

As mentioned in the above subsection, in the case of the point P3, the limit $(\Omm,\Ome)\rightarrow(0,0)$ must be interpreted asymptotically (i.e., when $L\rightarrow 0$). Given the scaling laws ruling the evolution of $\rho_{mq}$ and $\rho_{eq}$ for this coupling with the terms proportional to $L^{\gamma_2}$ being dominant over the terms with $L^{\gamma_1}$ as $\gamma_1>\gamma_2$,

\be
lim_{L\rightarrow 0}\frac{\Ome}{\Omm}=\frac{(\rho_{eq0}(1-a)+\rho_{mq0}b)L^{\gamma_2}}{(\rho_{mq0}a-\rho_{eq0}b)L^{\gamma_2}}=\frac{1-\Delta}{1+\Delta}
\ee
and, consequently, the direction of the limit near P3 is $k=(1-\Delta)/(1+\Delta)<1$, independently of the shell $r=r_i$ considered.
\begin{itemize}
\item P1: When the homogeneous part takes the form of the critical point P1, we find four different critical points:
    \begin{itemize}
    \item \underline{Saddle Point P1a}: $\Dh=0,\Db \textrm{arbitrary}, \Dm=0, \De=0$. While four of the corresponding eigen--values are  identical to these of the P1 that are listed in table \ref{table1} (with two of them equal to $\lambda_3 $, we find a null eigen--value (with eigen--vector in the direction of the $\Db$ axis) and, finally, the eigen--value $\lambda=(-3/2)(1+w(1-\Delta)/2)$. This point is a saddle point for the ranges of FPs considered.

    \item \underline{Points P1b,P1c,P1d}: where $\Dh$ is one of the roots of the polynomial
    \ba
    12&&\,{\Dh}^{3}+ a_1 {\Dh}^{2}+a_2 \Dh+a_3=0,\\
    a_1&&=3\,w\Delta-24\,\alpha-15\,w+2\\
    a_2&&= -6\Delta\alpha\,w+4\,w\Delta+12\,\alpha\,w-6\,w-2 \\
    a_3&&=(1-\,\Delta)(3{w}^{3}+4{w}^{2}+6\alpha w)+6\alpha{w}^{2}+4\alpha+2w
    \ea
    and $\Db=-1$, $\Dm=-1+2\alpha(\Dh-w)/((\Delta+1)(\Dh-2\alpha-w))$, $\De=-1-(w^2(\Delta-1)+w(\Delta+3)\Dh)/(2(\Dh-2\alpha-w))$. The eigen--values should be determined in a case by case basis. For example, when $w=-1$ and $\alpha=0.1$, only two of the critical points are physical and both are saddle points.

    \end{itemize}

\item P2: Assuming the homogeneous critical point P2, we find different choices of the $\delta$ functions that make the right hand side of eq. (\ref{evsistemdinamic3d}--\ref{evsistemdinamic3g}) null. One of them is a future attractor:
    \begin{itemize}
    \item \underline{Future attractor P2a}: $\Dh=0,\Db \textrm{arbitrary}, \Dm=0, \De=0$. In this case, the corresponding eigen--values are negative defined for any choice of the FPs: four identical to these of the P2 that are listed in table \ref{table1}, a null eigen--value (with eigen--vector in the direction of the $\Db$ axis) and, finally, the eigen--value $\lambda=(-3/2)(1+w(1+\Delta)/2)$, which is negative for the choices of $w$ and $\alpha$ considered. We conclude that this point act as a future attractor. As this point represents a future asymptotic point (when $L\rightarrow \infty$) for which the baryonic matter has a much lower contribution than the dark sources, the value of $\Db$ is irrelevant. For the dark sources, both $\delta$ functions tend to null, which correspond to a homogeneous metric.

    \item \underline{Points P2b, P2c,P2d}: where $\Dh$ is one of the solutions to
    \ba
    12&&\,{\Dh}^{3}+ b_1 {\Dh}^{2}+b_2 \Dh+b_3=0,\\
    b_1&&=-3\,w\Delta-24\,\alpha-15\,w+2\\
    b_2&&= 6\Delta\alpha\,w-4\,\Delta w+12\,\alpha\,w-6\,w-2 \\
    b_3&&=(1+\,\Delta)(3{w}^{3}+4{w}^{2}+6\alpha w)+6\alpha{w}^{2}+4\alpha+2w
    \ea
    and $\Db=-1$, $\Dm=-1+2\alpha(\Dh-w)/((1-\Delta)(\Dh-2\alpha-w))$, $\De=-1-(w^2(\Delta-1)+w(\Delta+3)\Dh)/(2(\Dh-2\alpha-w))$. This points should be treated similarly to the points P1b-P1c, in a case by case basis, e.g., when $w=-1$ and $\alpha=0.1$, only two of them are physical and they both behave as saddle points.

    \end{itemize}

\item P3: In this case, $k=(1-\Delta)/(1+\Delta)$ and we find three critical points of the complete system: two saddle points and an past attractor:
    \begin{itemize}
    \item \underline{Saddle point P3a}: $\Dh=0,\Db=0, \Dm \textrm{arbitrary}, \De=\Dm$. In this case, the diverging terms on eq. (\ref{evsistemdinamic3f},\ref{evsistemdinamic3g}) cancel and the corresponding eigen--values can be computed independently on the value of $k$. Four eigen--values take the same form as these in table \ref{table1} for P3 (with two of them identical to $\lambda_3$), one eigen--value is null, and the remaining are $-3/2$,and $3\alpha\frac{1-k^2}{k}>0$, respectively. This point behave as a saddle point.

    \item \underline{Past attractor P3b}: $\Dh=-1/2,\Db=-1$, $\De=\Dm$ and
    \be
    \Dm=-1-{\frac{2{\alpha}^{2}({k}^{2}+1)+2\alpha{k}^{2}(w+1)+4{\alpha}^{2}k+\alpha\,k}{2\alpha{k}^{2}-2\alpha-k}},\label{delmP3b}
    \ee
    In this case, we find two eigen--values that are defined as $3/2$, another one is $5/2$, three are defined as $\lambda_1$, $\lambda_2$ and $\lambda_3$ from P3 in table \ref{table1}, and finally, the last eigen--value is $-3w\Delta+3/2>0$. As all of them are positive defined, we conclude that P3b is a past attractor. 

    \item \underline{Saddle Point P3c}: $\Dh=1/3,\Db=-1$, $\De=\Dm$ and
    \be
    \Dm=-1-{\frac{3{\alpha}^{2}({k}^{2}+1)+3\alpha{k}^{2}(w-1)+6{\alpha}^{2}k-\alpha k}{3\alpha{k}^{2}-3\alpha-k}},
    \ee
    Two eigen--values are equal to $-1$ independently on the FPs, while the rest of them depend on $w$ and $\alpha$. For the range of parameters considered, we find at least two of the latter positive. Consequently, this is a saddle point.

    \end{itemize}

\item P4 Finally, for $P4$, we find two different critical points that are saddle points:
    \begin{itemize}
    \item \underline{Saddle point P4a}: $\Dh=0,\Db \textrm{arbitrary}, \Dm \textrm{arbitrary}, \De=\Dm$. One eigen--value take the form of $\lambda_1$ for P4 in table \ref{table1}, one is $\lambda_2$, while two of them are $\lambda_3=-1$, two are null and the remaining eigen--value depends on the direction $k$.  This point behave as a saddle point.

    \item \underline{Saddle Point P4b}: $\Dh=-1/3,\Db=-1$, and
     \ba
    \Dm&=&-1-{\frac{3{\alpha}^{2}({k}^{2}+1)+3\alpha{k}^{2}(w+1)+6{\alpha}^{2}k+\alpha k}{3\alpha{k}^{2}-3\alpha-k}},\\
    \De&=&-1-{\frac{3{\alpha}^{2}({k}^{2}+1)+3\alpha{k}^{2}w+6{\alpha}^{2}k-\alpha (k+1)+k w}{3\alpha{k}^{2}-3\alpha-k}}.
    \ea
    In this case, two eigen--values correspond to $\lambda_1$ and $\lambda_2$ for P4 in table \ref{table1}, respectively, two eigen--values are $\lambda_3=-1$, while we find two eigen--values with values $+1$, and two that depend on the choice of the FPs and the constant $k$.  We conclude that this is a saddle point.

    \end{itemize}

\end{itemize}

The past attractor point P3b and the future attractor P2a are of particular interest. In the next section, we define an initial profile example for which some computed trajectories evolve asymptotically from P3b to P2a.

\section{Initial profile leading to a structure formation example.}\label{numerical}

For some choices of initial profiles, we find a special kind of evolution for which some shells in the vicinity of the symmetry center (inner shells) $\HH_q =0$ at a finite time $t=t_{\hbox{\tiny{max}}}(r_i)$, the maximum expansion instant, while, for the rest of the shells (outer shells), $\HH_q$ evolve smoothly \cite{izsuss17a, izsuss17b}. For the inner shells, the QL curvature $\KK_q$ is necessarily positive (${\Omega}_{\KK}<0$) and the turning point is defined as
\be
\KK_q(r_i)=\frac{\kappa}{3}\,\left[\rho_{bq}+\rho_{mq}+\rho_{eq}\right]_{r_i,t_{\hbox{\tiny{max}}}}.
\ee
At the maximum expansion instant, the corresponding $\Omega$ scalars of the inner shells diverge while their corresponding QL energy--densities are finite. The evolution of the inner shells right after the maximum expansion instant cannot be described with the same LTB metric, but it is possible to match a contracting LTB solution for the inner shells with QL scalars defined by continuity, and, thus, obtain a spherically symmetric structure formation toy model. In terms of the dynamical system eqs. (\ref{evsistemdinamic3a}-\ref{evsistemdinamic3g}), a solution with divergent $\Omb$, $\Omm$, and $\Ome$ is not enough to define the spherical collapse toy model as some shell--crossing singularities can be present.

In this section, proceeding as in \cite{izsuss17a, izsuss17b}, we set the necessary conditions to avoid well-known singularities of the LTB metric, and we next define a set of local densities profiles and spatial curvature in order to numerically solve the evolution in $\xi$ at a fixed shell $r=r_i$. For the radial coordinate $r$, we consider $r \in [0,r_{\hbox{\tiny{max}}}]$ while $r_i$ is the element of an $n$-partition (defined as $r_i=i r_{\hbox{\tiny{max}}}/n$). The initial conditions are defined at the hypersurface $t=t_0$ (subindex $0$ is used for the different scalar evaluated at the instant $t_0$).

\subsection{Shell--cross singularities.}
The QL scalar formalism cannot be used at shell crossing singularities (as the delta functions are divergent). Thus, the initial conditions considered in the numerical work must avoid an evolution towards a shell crossing, i.e., $\Gamma>0$ must hold throughout the evolution. For LTB dust solutions with $\Lambda=0$ it is possible to state analytic restrictions on initial conditions to guarantee an evolution with no shell crossings (\cite{ltbrev1,ltbrev2,ltbrev3,RadProfs}), but for the solutions with nonzero pressure that we are considering here this can only be achieved by numerical trial and error of initial conditions.

In order to integrate the system (\ref{evsistemdinamic3a})--(\ref{evsistemdinamic3g}), we can compute the initial conditions from a given set of initial profiles $\rho_{b0}(r), \rho_{e0}(r), \rho_{m0}(r)$ and $\KK_{0}(r)$ and a defined $R_0(r)$.  The QL scalars initial profiles $\rho_{bq0}(r),\,\rho_{eq0}(r),\, \rho_{mq0},\,\KK_{q0}(r)$ and the fluctuations $\Dm_0,\,\De_0,\,\Dh_0$ follow directly from (\ref{qmaps}) with $R=R_0$. For simplicity, it is convenient to consider  $\xi_0=\xi(t_0,r)=0$ for all $r$\footnote{It is possible to set $\xi=0$ to define the same hypersurface as $t=t_0$, and from this point on, $t$ and $\xi$ hypersuperfaces will differ.}, which leads to:  $\Ombi=\Omb(0,r)$, $\Omei=\Ome(0,r)$, $\Ommi=\Omm(0,r)$, $\Db_0=\Db(0,r)$, $\Dm_0=\Dm(0,r)$, $\De_0=\De(0,r)$ and $\Dh_0=\Dh(0,r)$.

The initial profiles could evolve to a shell crossing singularity, so we carefully test the evolution of $\Gamma$ to verify that $\Gamma>0$ holds for all $\xi$. It is convenient to relate $\Gamma$ to $\Dk$ by deriving with respect to radial coordinate the scaling laws (\ref{scalinglawrhomq})
\be
\Dk=-\frac{2}{3}+\frac{\frac{2}{3}+\Dk_0}{\Gamma}\quad \Rightarrow \quad \Gamma=\frac{\frac{2}{3}+\Dk_0}{\frac{2}{3}+\Dk},
\label{Deslaw}
\ee
which we can rewrite, from  (\ref{cHam3b}), as
\be\label{gamma}
\fl \Gamma=\frac{{\Omega}_{\KK}(t,r)}{{\Omega}_{\KK}(0,r)}
\left(\frac{\Ombi(\Db_0+\frac{2}{3})+\Ommi(\Dm_0+\frac{2}{3})+\Omei(\De_0+\frac{2}{3})-2\Dh_0-\frac{2}{3}}{\Omb(\Db+\frac{2}{3})+\Omm(\Dm+\frac{2}{3})+\Ome(\De+\frac{2}{3})-2\Dh-\frac{2}{3}}\right).
\ee
From the expression of above, it is clear that $\Gamma$ tends to null as $\Db$, $\Dm$, $\De$ diverge. On the other hand, $\Gamma$ do not mandatorily diverge as $\HH_q$ tends to zero. In the later case, the $\Omega$ functions and $\Dh$ in the denominator will diverge, but the factor ${\Omega}_{\KK}(t,r)$ will diverge as well at the same rate as the rest of the $\Omega$ functions and faster than $\Dh$ ($\Dh$ and the $\Omega$ functions inversely proportional to $\HH_q$ and to $\HH_q^2$, respectively). In general, an evolution of the models that is free from shell crossings needs to be determined numerically from (\ref{gamma}).

\subsection{Initial profiles, evolution and collapse time}

As mentioned in \cite{izsuss17a}, the LTB metric has a scale invariance that allows us to define dimensionless quantities:  cosmic time $\bar t=H_0 t$,  Hubble factor $\bar \HH_q=\HH_q/H_0$, and local densities $\kappa\bar\rho_{a}/3= \kappa\rho_a/(3H_0^2)$ with $H_0$ an arbitrary constant (typically chosen as the present day Hubble factor in units $Km/(Mpc\cdot s)$ in cosmology) and subindex $a=b,m,e$. For the numerical work in this section, we take the time units with $H_0=1$ and energy density units with $\kappa/(3H_0^2)=1$.

We consider the initial local profiles
\ba
\rho_{b0}&=&{ b_{10}}+{\frac {{ b_{11}}-{ b_{10}}}{1+\tan^{2}(r)}},\qquad { b_{10}}= 0.90,\quad { b_{11}}= 1.20;\nonumber\\
\rho_{m0}&=&{ m_{10}}+{\frac {{ m_{11}}-{ m_{10}}}{1+\tan^{2}(r)}},\qquad { m_{10}}= 1.00,\quad { m_{11}}= 13.10;\nonumber\\
\rho_{e0} &=&{ e_{10}}+{\frac {{ e_{11}}-{ e_{10}}}{1+\tan^{2}(r)}},\qquad { e_{10}}= 0.90,\quad { e_{11}}= 1.66;\\
\KK_0&=&k_{10}+\frac{k_{11}-k_{10}}{1+\tan^2(r)},\qquad k_{10}=-1.10, \quad k_{11}=2.50;\nonumber
\label{struc}
\ea
and $R_0(r)=\tan(r)$, with an $r$ coordinate partition has $n=20$ elements from $0$ to $r_{\hbox{\tiny{max}}}=\pi/2$. This choice is specially useful for the numerical work, as it would be possible to define a physical $\mathfrak{R}$ coordinate from $R_{0}$ and $r$ as $\mathfrak{R}=\tan(r)$ for which $\mathfrak{R}=0$ correspond to $r=0$ (the symmetry center) and $\mathfrak{R} \rightarrow \infty$ as $r\rightarrow r_{\hbox{\tiny{max}}}$. For the FPs, we chose $w=-1$,  as in this case DE represents a well known cosmological constant.  Although the observational bounds in \cite{ol,ol2,ol3, wang16} on the parameter $\alpha$ for this interaction suggest a value close to zero (of order $\sim 0.001$), we take $\alpha=0.1$ in order to stress the effect of the coupling term on the dynamics.

The initial profiles of above evolve to a LTB collapse scenario as some shells with values of $r=r_j$ around the symmetry centre $r=0$ initially expand ($\HH_q(t,r_j)>0$), then bounce as $\HH_q(t=t_{\hbox{\tiny{max}}},r_j)=0$ and, finally, collapse ($\HH_q$ negative), whereas the rest shells expand. For each bouncing shell with $r=r_j$, it is possible to numerically compute $t=t_{\hbox{\tiny{max}}}$. As the inner shells evolve to the bounce instant, $\Omb,\,\Ome,\,\Omm \rightarrow \infty$. In this numerical example, we do not address the collapsing stage of evolution of the inner shells.

\begin{figure}[tbp]
\includegraphics*[scale=0.50]{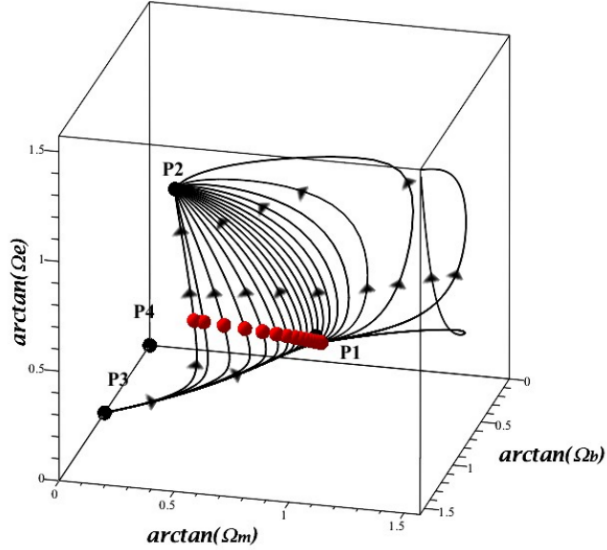}

\caption{Phase-space trajectories of the homogeneous subsystem (\ref{evsistemdinamic3a}--\ref{evsistemdinamic3c}) with initial conditions (\ref{struc}) and FP $w=-1.0$, $\alpha=0.1$. The red dots represent the initial values of $\Omb$, $\Omm$ and $\Ome$ at every shell $r=r_i$ of the partition considered. The black points represent the critical points of the homogeneous projection from table \ref{table1}.} \label{fig1}\end{figure}

Figure \ref{fig1} depicts the evolution of $\Omb$, $\Omm$ and $\Ome$ for the different shells of the partition of $r$ plotted in the phase--space of the homogeneous subsystem. We have plotted $\arctan(\Omega)$ in order to obtain a finite value of this function when $\Omega\rightarrow \infty$ (as $\Omega\rightarrow \infty$ implies $\arctan(\Omega)\rightarrow \pi/2$). Red dots represent the initial conditions for every shell $r=r_i$ in the radial partition. From the numerical results we notice that the first and second shells $j=1,2$ (close to the symmetry center) evolve from the past attractor $P3$ to infinity (corresponding to the instant when $\HH_q=0$ if no shell crossing singularities occur), while the rest of shells evolve from the past attractor $P3$ to the future attractor $P2$. Also, the trajectories evolve near the vicinity of the saddle point $P1$, stressing its behavior as an attractor in the direction of the eigenvector parallel to $\Omb$ (related to the corresponding eigenvalue $\lambda_2$ in \ref{table1}). This attractor is unstable in the plane generated by the eigenvectors related to $\lambda_1,\, \lambda_2$ in \ref{table1}. The later unstable plane is parallel to the plane $\Omb=0$. This numerical example representation in the homogeneous projection allows us to have a better understanding of the critical points of the homogeneous projection.

\begin{figure}[tbp]
\includegraphics*[scale=0.30]{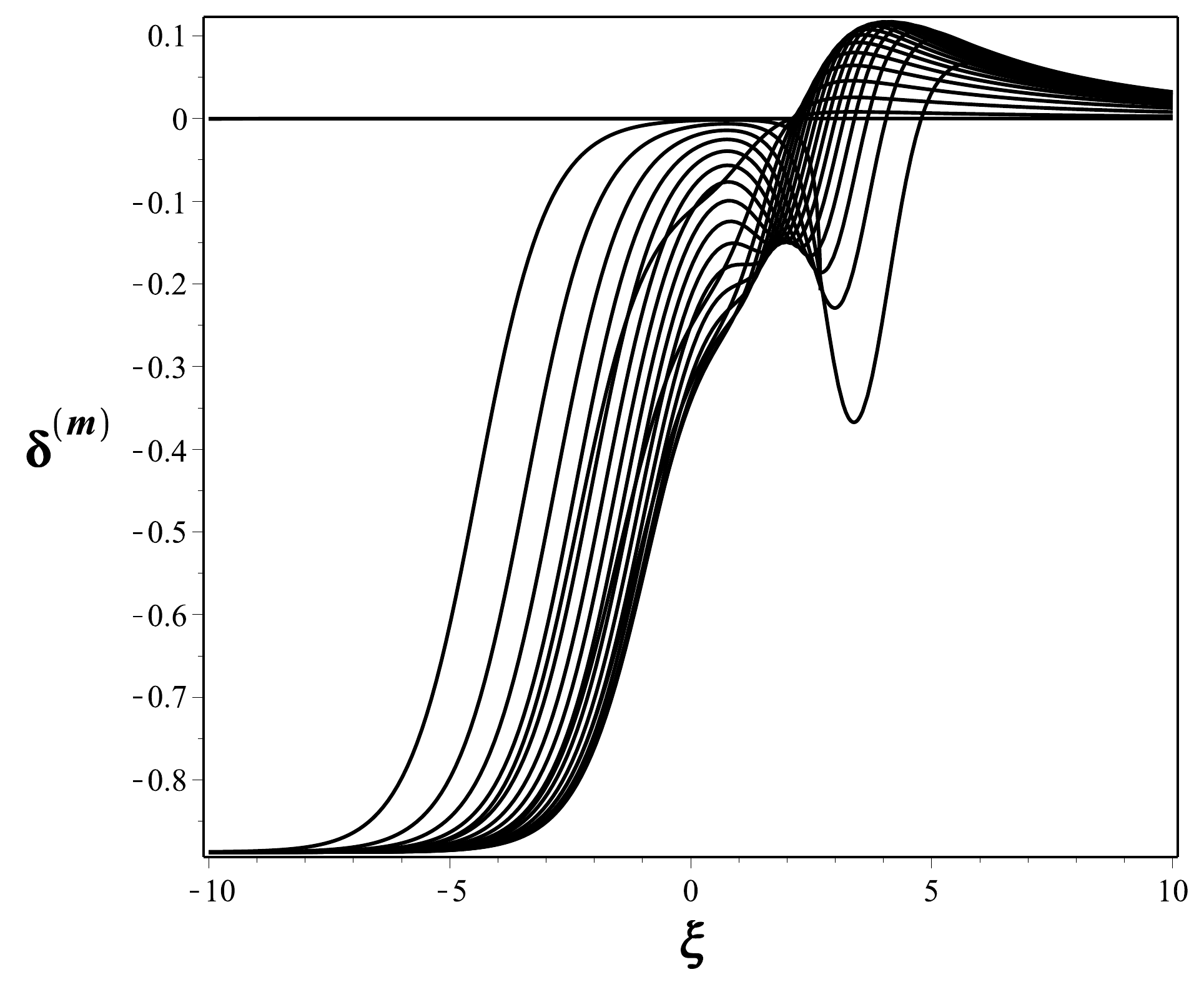}
\includegraphics*[scale=0.31]{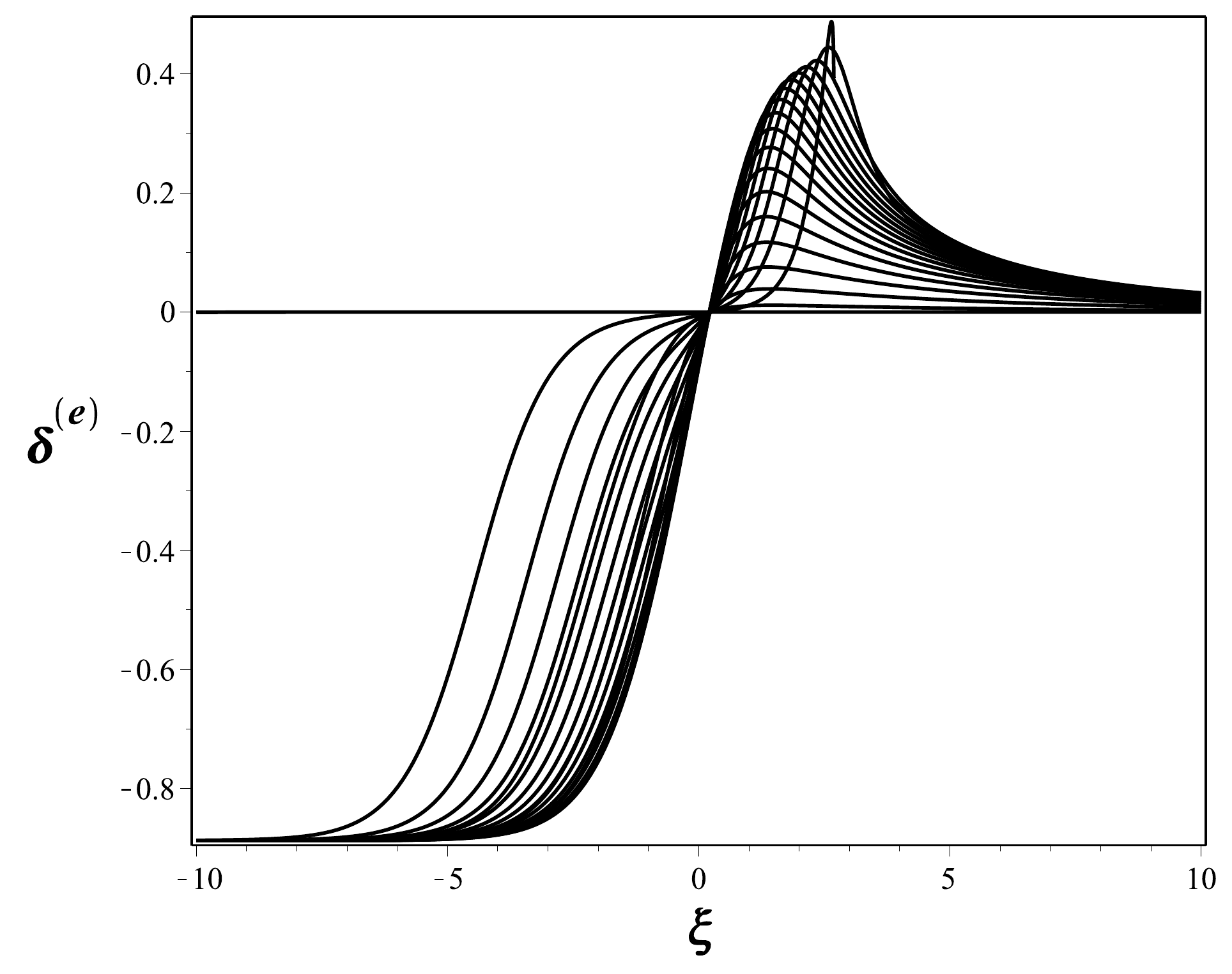}
\includegraphics*[scale=0.31]{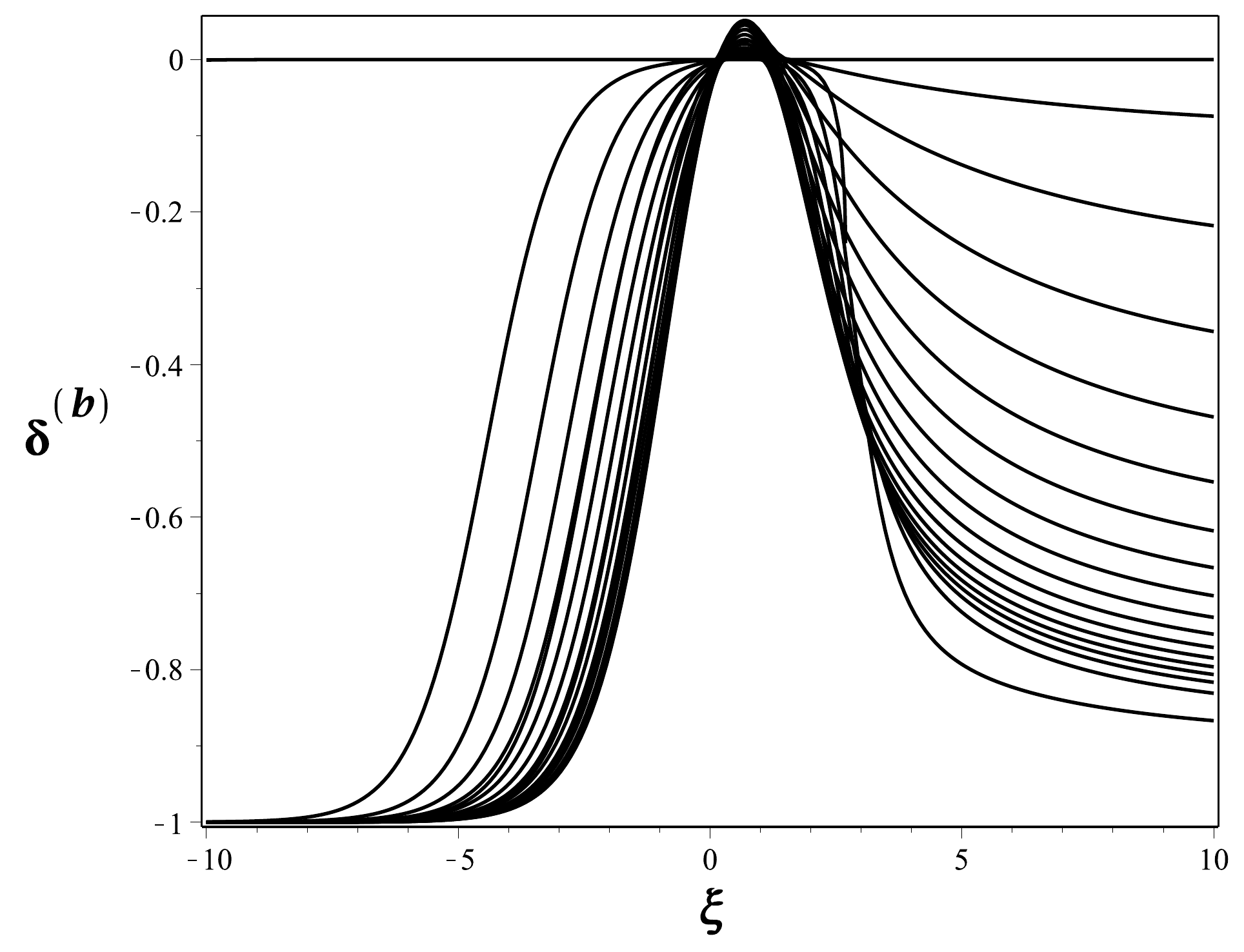}
\includegraphics*[scale=0.31]{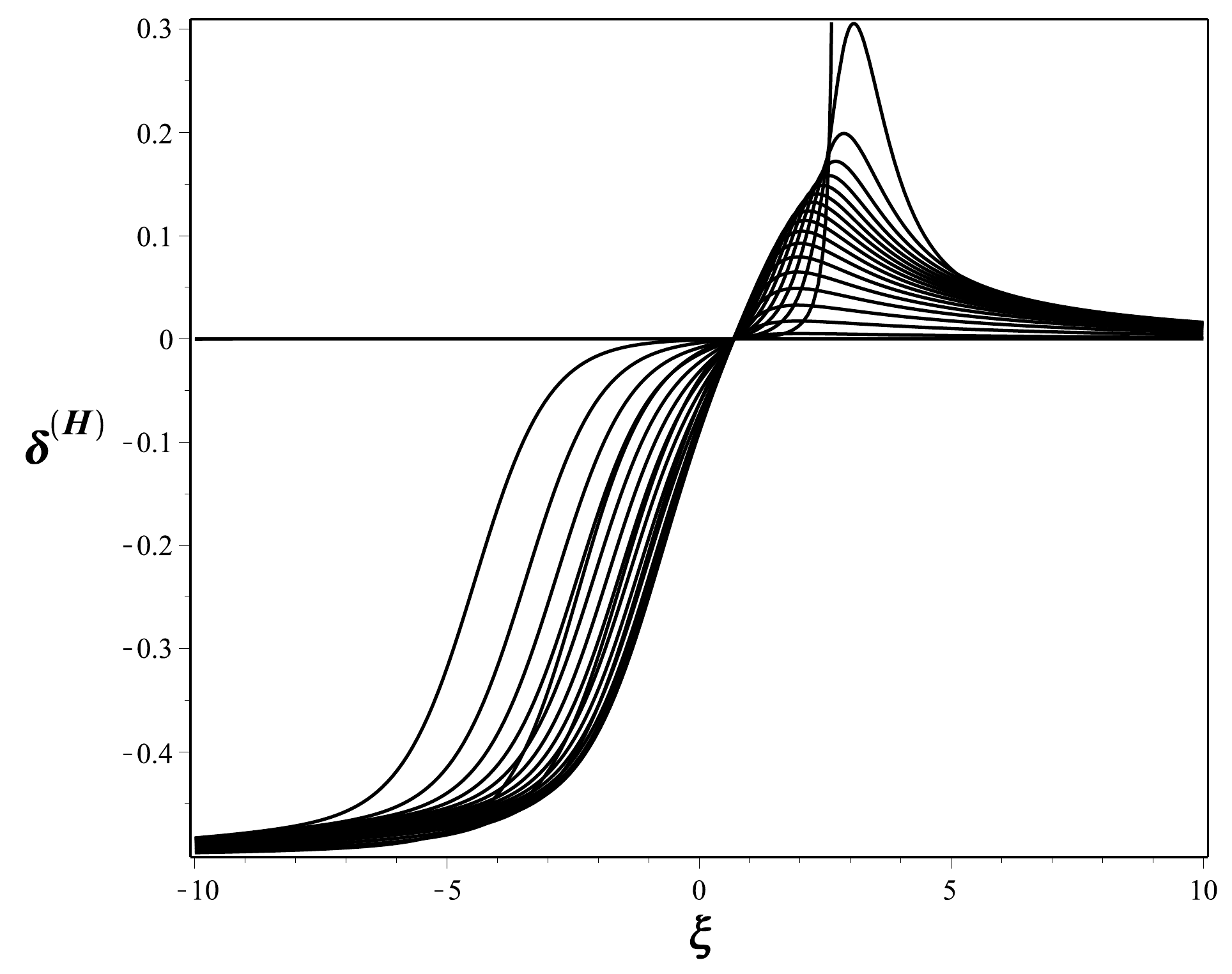}
\caption{Evolution of $\Dm$, $\De$, $\Db$ and $\Dh$ vs. $\xi$ for shells $r=r_i$ with initial conditions given by (\ref{struc}) and $w=-1.0$, $\alpha=0.1$. For all the shells, the $\delta$ functions evolve from the saddle point P3b in the past (negative values of $\xi$). For the outer shells, the $\delta$ functions evolve to the future attractor P2a.} \label{fig2}\end{figure}

In order to check for possible shell crossing singularities for this profile, it is necessary to compute evolution of $\Dm$, $\De$, $\Db$ and $\Dh$ for all the shells. As discussed above, if $\Dm$, $\De$, $\Db$ tend to infinity if $\Gamma\to 0$ that marks a shell crossing singularity. As shown in figure \ref{fig2}, the values of $\Dm$, $\De$, $\Db$ for all $\xi$ remain bounded and thus no shell crossing singularities occur for the initial profile. The function $\Dh$ diverges as $\HH_q$ tend to null for the inner shells, as expected from its definition $\Dh=\HH/\HH_q-1$, but, as discussed previously, $\Gamma$ do not tend to null for those shields as the factor $\Omega_{\KK}$ in (\ref{gamma}) is inversely proportional to $\HH_q^2$ and, consequently, $\Dh/\Omega_{\KK} \rightarrow 0$ as $\HH_q \rightarrow 0$.

In figure \ref{fig2}, we see how the $\delta$ functions of the outer shells evolve towards the critical point P2a. For the larger $\xi$ values, $\Dh$, $\Dm$ and $\De$ tend to null for all the shells, and $\Db$ tend to an arbitrary value. In this numerical example and in the limit $L\rightarrow0$, the ratio $\Omm/\Ome$ tends to the constant $k=0.1270$, which leads to the positive defined eigen--values of P3b, i.e., in this example, P3b acts as a past attractor for all the shells. Consequently, we find that the $\delta$ functions evolve in the past to the values found in the point P3b: $\Dh$ tends to $-1/2$ in the past, $\Db$ tends to $-1$, while $\Dm$ and $\De$ tend to the corresponding values of eqs. (\ref{delmP3b}, \ref{delmP3b}), which is $-0.8873$ for both functions.

\begin{figure}[tbp]
\includegraphics*[scale=0.50]{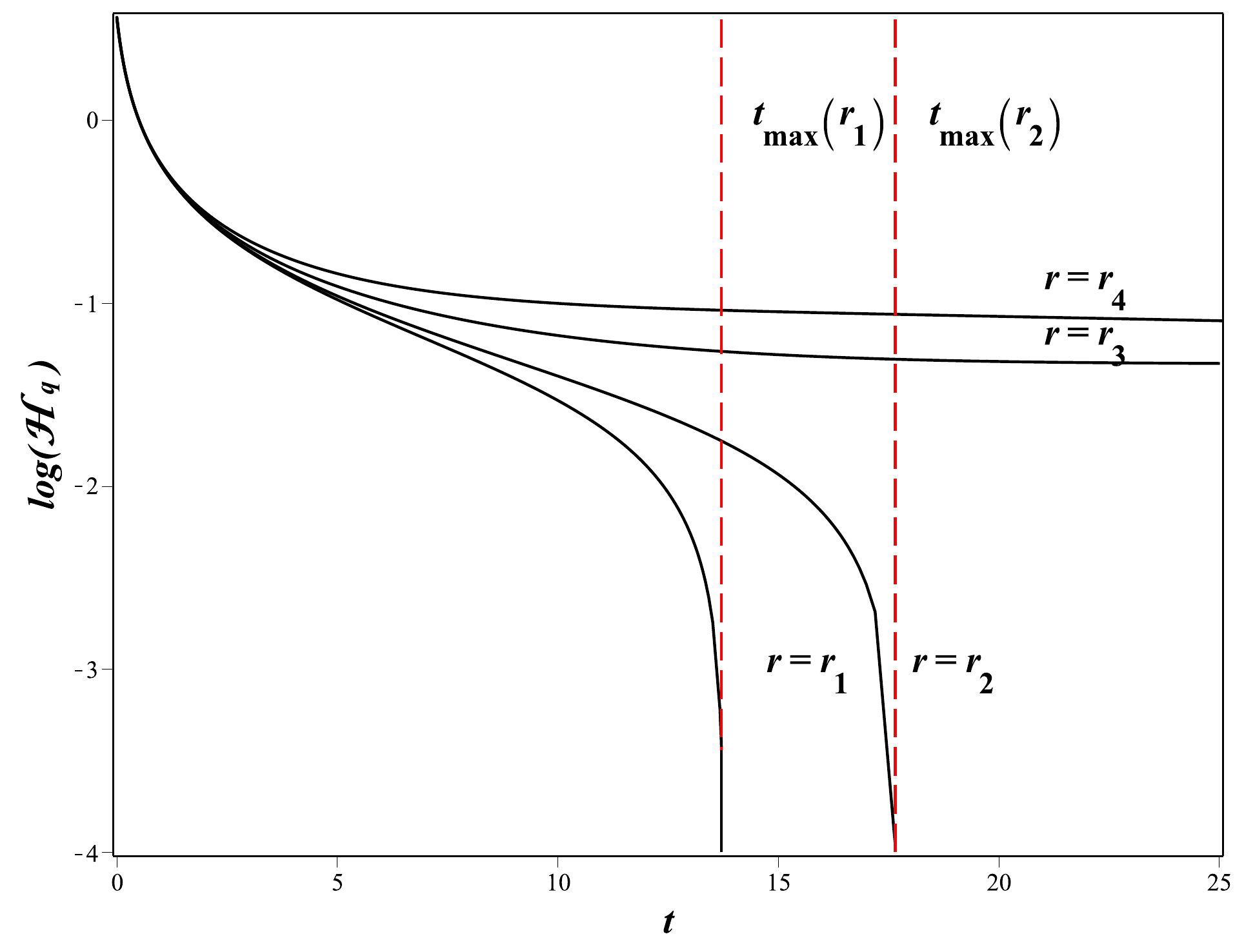}

\caption{Evolution of $\log(\HH_q)$ (from \ref{HOm}) vs. cosmic time $t$ (defined in (\ref{phystime})) for shells $r_1$, $r_2$, $r_3$, and $r_4$ with initial conditions given by (\ref{struc}) and $w=-1.0$, $\alpha=0.1$. While $\HH_q$ of the inner shells $r_1$, $r_2$ tend to null, $\log(\HH_q)$ tends asymptotically to $-\infty$. For outer shells $r_3$, $r_4$, $\HH_q$ is a continuous function, so is $\log(\HH_q)$.} \label{fig3}\end{figure}

With the numerical solution of the dynamical system, we can compute  $\HH_q$ from (\ref{HOm}), with the cosmic time $t$ defined in (\ref{phystime}). We can, then, plot implicitly the evolution of $\log(\HH_q)$ vs. $t$ for every shell $r=r_i$ of the partition. For the bouncing inner shells ($r_1$ and $r_2$ in this case), it is possible to evaluate numerically the cosmic time for which $\HH_q=0$ (asymptotic behaviour in $\log(\HH_q)$): $t_{\hbox{\tiny{max}}}(r_1)=13.71$ and $t_{\hbox{\tiny{max}}}(r_2)=17.65$, respectively. Note that those cosmic times correspond to the maximum expansion of both shells. In figure \ref{fig3}, the curves $\log(\HH_q)$ vs. $t$ correspond to the shells $r_1$, $r_2$, $r_3$, and $r_4$.

The instant $t_{\hbox{\tiny{max}}}(r_j)$ ($j=1, 2$ for the numerical example of above) in our LTB model plays a role analogous to the turnaround time in the Newtonian spherical collapse model \cite{pad} or the  collapse of a spherical perturbation in a FLRW background (see \cite{gunn72} and \cite{mo10} for a top hat profile spherical collapse in an Einstein--de Sitter and a $\Lambda$CDM background, respectively), which in our scenario corresponds to the shells $r=r_j$ reaching their maximal expansion at different times $t=t_{\hbox{\tiny{max}}}(r_i)$. Therefore, the numerical example we are presenting represents a collection of ``bowler hat'' profiles (smoothed ``top hats'') in which two shells of the partition collapse (with different values for $ t_{\hbox{\tiny{max}}}(r_i)$). It is reasonable then to average the values $t_{\hbox{\tiny{max}}}(r_i)$ to obtain a single turnaround instant given by: $\langle t_{\hbox{\tiny{max}}}\rangle =15.68$.

Note that the average $\langle t_{\hbox{\tiny{max}}}\rangle$ is obtained numerically, hence it depends, not only on the background dynamics: FPs $w$ and $\alpha$, energy densities/expansion rate of the background (as is the case in the collapse of spherical perturbations in a FLRW background), but also on the chosen local initial profiles. While a single turn around value $\langle t_{\hbox{\tiny{max}}}\rangle$ can always be found, it is necessary to do it in a case by case basis on the full non-linear dynamics (as opposed to a linear order approximation in the spherical collapse model).

\section{Interaction proportional to CDM and DE energy densities vs. other similar coupling terms} \label{Interaction}

It is important to compare the results of the present article with those of previous work on the dynamics of LTB models described by evolution equations for the QL scalars for a different mixtures of coupled CDM and DE sources. In \cite{izsuss17b}, the mixture that was considered consisted in two dark fluids coupled by an interaction term $J_{q}=3\alpha\HH_{q} \rho_{mq}$, which lead to a 5--dimensional autonomous dynamical system with the homogeneous projection defined by the functions $\Omm$ and $\Ome$ and the inhomogeneous projection defined by $\Dm$, $\De$ and $\Dh$. In \cite{izsuss17a},  we considered the same sources and a coupling term given by $J_{q}=3\alpha \HH_{q} \rho_{eq}$, with the addition of a radiation fluid in the homogeneous projection (dominant near the initial singularity) examined in an appendix.

In this work, we consider an additional non relativistic matter term (baryonic matter) which is consistent with the cosmological observations and the coupling term is more general:  $J_{q}=3\alpha \HH_{q} (\rho_{mq}+ \rho_{eq})$. As a consequence, the corresponding homogeneous projection is now 3--dimensional ($\Omb$, $\Omm$ and $\Ome$) and the inhomogeneous projection has an additional function $\Db$. The addition of the baryonic matter source makes the dynamical system study more complex, specially the inhomogeneous projection.

We can compare the homogeneous projection in this work with those of previous papers, as some similarities arise. On one hand, we see that the homogeneous projections in the three articles contain a similar future attractor, whose position is determined by the type of interaction considered, as well as by the FPs $w$ and $\alpha$ (the future attractor has coordinates: $\Omm=0,\Ome=1$ in \cite{izsuss17b}; $\Omm=-\alpha/w,\Ome=1+\alpha/w$ in \cite{izsuss17a}; and $\Omb=0,\Omm=(1-\Delta)/2,\Ome=(1+\Delta)/2$ with $\Delta^2=1+4\alpha/w$ in this work). On the other hand, the past attractor in both \cite{izsuss17b} and \cite{izsuss17a} (with coordinates $\Omm=1+\alpha/w, \Ome=-\alpha/w$ and $\Omm=1,\Ome=0$ respectively) can be easily related to the critical point $P2$ in the present work ( $\Omb=0,\Omm=(1+\Delta)/2,\Ome=(1-\Delta)/2$). Note that, while $P2$ is a saddle point, it behaves as a past attractor in the subspace $\Omm - \Ome$ (generated by the eigenvectors with positive defined eigenvalues), while it acts as an attractor in the direction parallel to $\Omb$ axis. We can conclude that the past attractor (whose position is dependent of the interaction and FP considered) of the previous papers changed to a saddle point when we add an extra baryonic matter source. In fact a new past attractor $P3$ is added in this work, which can play an interesting role in the evolution of the initial profiles.

An invariant line of the homogeneous projection was found in \cite{izsuss17a}, connecting the origin of the $\Omm-\Ome$ plane with the future attractor. In \cite{izsuss17b}, we found also an invariant line connecting $\Omm=0,\Ome=0$ point and the past attractor. Given that no phase space trajectory can cross this later invariant line, some initial conditions, near the past attractor (and under this line) would end their evolution towards the $\Ome=0$ axis (see figure 1 in \cite{izsuss17b}). Looking at the projection $\Omb=0$ in the present paper, we see that neither the line that connects the origin with the saddle point $P2$ nor the line that connects the origin with the future attractor are invariant subspaces. Although there can still be trajectories that evolve to $\Ome=0$ axis in this work, they are related to the $\Omb=0$ projection and can be avoided in a cosmological scenario by considering an extra baryonic matter source.

For the interaction $J_{q}=3\alpha \HH_{q} (\rho_{mq}+ \rho_{eq})$ in the 3-dimensional homogeneous projection, we find the invariant plane $\Omega_{\KK}=-\KK_q/\HH_q^2=0$. In fact $\Omega_{\KK}=0$ is also an invariant subspace in the previous works, defined as the lines that connect the past and future attractors. In all the cases, the trajectories of the homogeneous phase--space can be classified according to their curvature sign: flat trajectories $\Omega_{\KK}=0$, where the shells evolve from the past attractor to the future attractor through the invariant line/plane; trajectories with $\Omega_{\KK}<0$, that evolve asymptotically from the past attractor to the future attractor according to their corresponding scaling laws maintaining the same sign in the curvature; and trajectories $\Omega_{\KK}>0$ that evolve asymptotically from the past attractor to the future attractor or to infinity (the later trajectories representing a collapse shell in the complete system description if no singularities are found).

In order to define the $\Gamma$ function in \cite{izsuss17b} and \cite{izsuss17a}, analytical solutions where studied for $\rho_{mq}$ and $\rho_{eq}$ in terms of the scale factor $L$. In \cite{izsuss17b} and because of the choice of the interaction, the function $\rho_{mq}$ was particularly simple as $\rho_{mq}=\rho_{mq0}L^{-3(1-\alpha)}$, allowing us to connect the $\Gamma$ function directly to $\Dm$. On the other hand, in \cite{izsuss17a}, it was  $\rho_{eq}$ that had a power law dependency on $L$ as $\rho_{eq}=\rho_{eq0}L^{-3(1+w+\alpha)}$, which resulted in $\Gamma$ being related to $\De$. In the present paper with a more general interaction and the addition of the baryonic matter source, we find an analytical solution for $\rho_{mq}$ and  $\rho_{eq}$ in terms of $L$ but no simple power law was obtained for either one of them, making the $\Gamma$ function dependent on $\Db$, $\Dm$ and $\De$.

Finally, it is worth considering the addition of a nearly homogeneous radiation-like source to the dynamical system, proceeding in a similar way as we did in the appendix in \cite{izsuss17a}. Such radiation source would yield a new function $\Omega_r$ that would be added to the set of phase space variables (with $\delta_r=0$, since it would be homogeneous radiation), thus leading to a 4-dimensional homogeneous projection. It is natural to assume that the critical point $P3$ (related to a model without CDM or DE sources) would no longer be an attractor of the system, as a new attractor would appear at the value $\Omega_r=1$, with the rest of the $\Omega$ functions vanishing. This attractor is easily understood given the evolution with the scale factor $L$ of the QL scalars. The QL scalars $\rho_{bq}$, $\rho_{mq}$ and $\rho_{eq}$ would still evolve with $L$ as in eq. (\ref{scalinglawrhomq}), while $\rho_{rq}=\rho_{rq0}L^{-4}$. As $L \rightarrow 0$ for every shell, the radiation source would dominate the early expansion near the initial singularity in the same way as in \cite{izsuss17a}.

As a concluding remark, the addition of the baryonic source and the consideration of a more general coupling term in this work added more complexity to the dynamical system analysis, but it is still possible to analyticaly obtain interesting information and a similar kind of numerical solutions as in previous articles.

\section{Conclusions} \label{conclusions}

We have generalized in the present paper the results of previous work \cite{izsuss17a, izsuss17b} by studying the dynamics of LTB solutions containing three sources: baryonic matter with interactively coupled CDM and DE. The coupling term we considered is a reasonable generalization of those used in the previous papers, as it is proportional to the addition of both dark sources energy densities (not only to one of them). Using the QL scalars formalism, we transformed the Einstein's evolution equations into a 7--dimensional autonomous dynamical system. The dynamical system  can be decomposed in two subsystems: the 3--dimensional invariant homogeneous subsystem whose variables are the dimensionless QL scalars ($\Omb,\,\Omm,\,\Ome$), related to the FLRW model, and a 4--dimensional subspace for the $\delta$ functions ($\Db, \Dm,\,\De,\,\Dh$) that can be interpreted as exact deviations from the FLRW background.

For the homogeneous projection we obtained four critical points summarized in table \ref{table1}: a saddle point $P1$, a future attractor $P2$, a past attractor $P3$ and a saddle point $P4$. The behavior of the critical points was examined under the assumption that $w$ is of order/ lower than $-1$ and $0<\alpha<0.25$, based on the observational bounds that have been obtained for this coupling in FLRW cosmology  \cite{ol,ol2, ol3, wang16}. In the complete description and given the complexity of the dynamical system, up to thirteen critical points are found (some of them not compatible with spherical symmetry for some choices of the FPs, with $\delta<-1$). Of particular interest are the past attractor $P3b$ and the future attractor $P2a$, that should be considered as asymptotical points in the limits $L\rightarrow 0$ and $L\rightarrow \infty$ for the expanding LTB metric ($L$ being a FLRW scale factor--like function), respectively. The future attractor shows a homogeneous LTB space where CDM and DE are dominant sources with null $\delta$ functions while curvature and baryonic matter have much smaller contributions. The $\delta$ functions are computed with a set of initial conditions in order to avoid shell crossing singularities (instant for which $\Gamma\to 0$, with $\Gamma$ defined by (\ref{gamma})). To have $\Gamma>0$ is a necessary condition to avoid shell crossing singularities for which $\Db, \Dm,\,\De,\,\Dh$ diverge at finite evolution times.

Finally, in order to illustrate how to solve numerically the evolution equations we considered a specific  example of a given set of initial profiles (\ref{struc}), with shell partition for the coordinate $r$ and FPs ($w=-1$ and $\alpha=0.1$). In this example, two inner shells ($r_1$ and $r_2$) evolved with $\Omb,\,\Omm,\,\Ome$ tending to infinity at a finite time while the outer shells evolve towards the future attractor (figure \ref{fig1}). As no shell crossing is present during the evolution (figure \ref{fig2}), we can conclude that the $\Omb,\,\Omm,\,\Ome$ diverging for the inner shells corresponds to a maximal expansion or ``turn around'' instant for which $\HH_q=0$ marks the outset of a collapse scenario. Because the LTB models are inhomogeneous these ``turn around'' instants occur at different cosmic times for different comoving observers,  though it is possible to characterize these times as a single ``turn around'' by computing numerically (in a case by case basis) the average $\langle t_{\hbox{\tiny{max}}}\rangle$ (figure \ref{fig3}). In this way we can relate this numerical computation to the single ``turn around'' time in ``top hat'' of spherical perturbations models discussed in the literature.

\end{document}